\documentclass[10pt,conference,letterpaper]{IEEEtran}
\IEEEoverridecommandlockouts
\usepackage[T1]{fontenc}
\usepackage{amsmath,amssymb}
\usepackage{graphicx}
\usepackage[caption=false,font=footnotesize,listofformat=parens]{subfig}
\usepackage{booktabs}
\usepackage{array}
\usepackage{cite}
\usepackage{url}
\usepackage{microtype}
\usepackage[colorlinks=false,pdfborder={0 0 0}]{hyperref}
\newcommand{\codeurl}[1]{%
  \begingroup
  \hypersetup{pdfborder={0 0 1},urlbordercolor={0 1 0}}%
  \url{#1}%
  \endgroup
}
\title{SiGMA: Site-Grounded Multi-Modal ISAC Data Generation for Environment-Aware Communications}
\author{
    \IEEEauthorblockN{
        Yijie Bian\IEEEauthorrefmark{1},
        Wei Guo\IEEEauthorrefmark{1},
        Kai Zhang\IEEEauthorrefmark{1},
        Xiao Ye\IEEEauthorrefmark{2},
        Shurui Xu\IEEEauthorrefmark{2},
        \\
        Shenghui Song\IEEEauthorrefmark{1}, \textit{Senior Member, IEEE},
        Jun Zhang\IEEEauthorrefmark{1}, \textit{Fellow, IEEE}, and
        Khaled B. Letaief\IEEEauthorrefmark{1}, \textit{Fellow, IEEE}
    }
    \IEEEauthorblockA{\IEEEauthorrefmark{1}Dept. of Electronic and Computer Engineering, The Hong Kong University of Science and Technology, Hong Kong, China}
    \IEEEauthorblockA{\IEEEauthorrefmark{2}School of Information Science and Engineering, Southeast University, Nanjing, China}
    \vspace{-0.45cm}
}
\begin{document}
\maketitle

\begin{abstract}
Learning-based multi-modal integrated sensing and communication (ISAC) requires data that associate environmental observations with the wireless response generated by the same site and dynamic scene state. Acquiring such data at new sites requires sensor calibration, coordinated measurements, and cross-modal association. We present site-grounded multi-modal ISAC data generation (SiGMA), which combines OpenStreetMap (OSM), Overture Maps, and procedural three-dimensional (3D) reconstruction to create georeferenced urban scenes. A shared scene geometry and time-indexed actor state link synchronized red-green-blue (RGB) images, light detection and ranging (LiDAR) point clouds, positioning records, frequency-modulated continuous-wave (FMCW) Radar in-phase and quadrature (IQ) samples, and ray-traced wireless channels. For configurable reflected propagation under direct-link blockage, SiGMA generates optional reconfigurable intelligent surface (RIS) component channels for phase-configuration evaluation across shared scene states. The evaluation results demonstrate that the framework improves zero-shot beam prediction performance on real DeepSense 6G measurements compared with an OSM-derived synthetic-data baseline. A camera-assisted RIS case study further demonstrates the generation of sensing-associated supervision for configurable reflected propagation. The dataset generation pipeline is available 
at \codeurl{https://github.com/Trans-cending/SiGMA-main}.
\end{abstract}

\begin{figure*}[t]
\centering
\includegraphics[width=.95\textwidth]{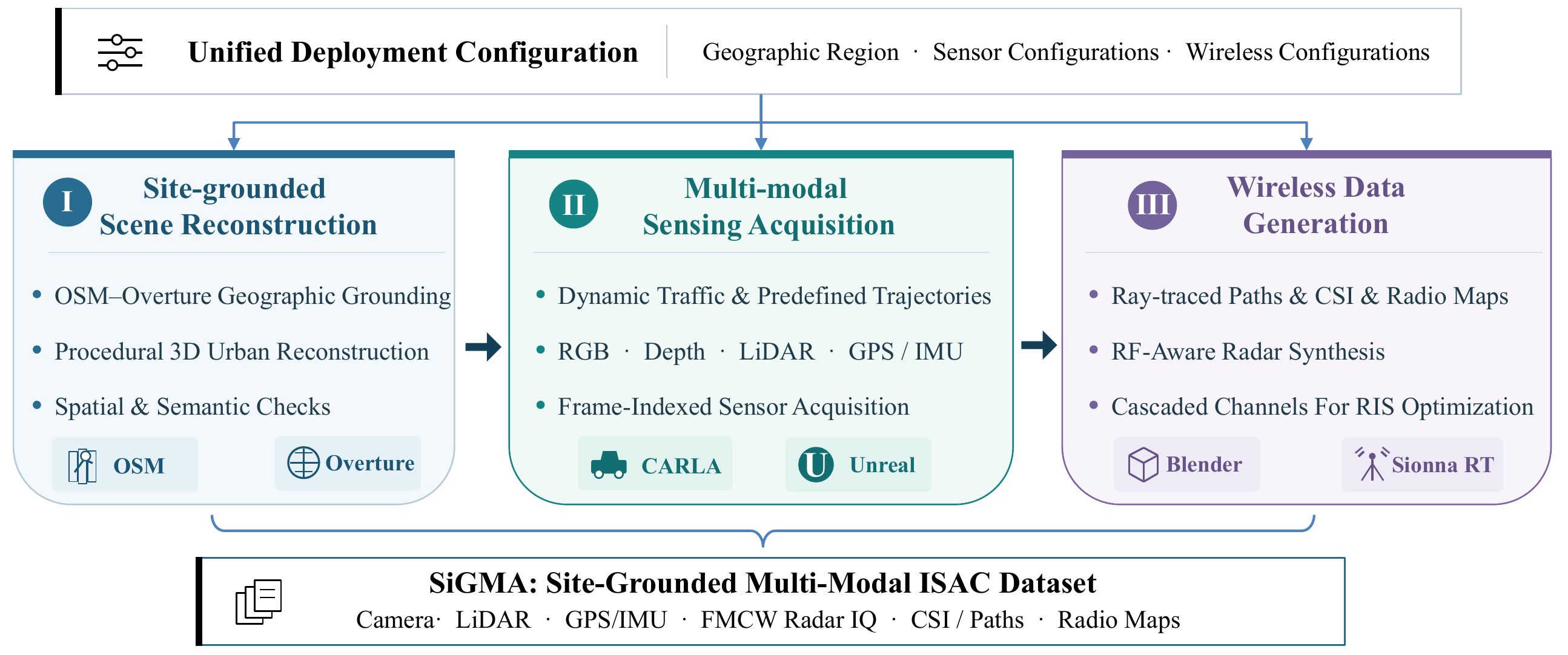}
\caption{SiGMA data-generation pipeline. Geographic reconciliation and procedural reconstruction establish a site-grounded scene for CARLA traffic and sensor acquisition. The corresponding scene states drive ray-traced wireless channels and radio detection and ranging (Radar) waveform synthesis.}
\label{fig:overview}
\end{figure*}

\section{Introduction}\label{sec:intro}
Sixth-generation (6G) systems are envisioned to integrate communication, environmental awareness, and native intelligence~\cite{letaief2019roadmap6g}. Integrated sensing and communication (ISAC) couples sensing and communication through shared radio resources and infrastructure~\cite{liu2022isac}. Environmental observations can inform wireless decisions, while radio signals can probe the scene. Buildings and other static obstacles determine direct-path visibility and available propagation paths, while moving objects change link blockage over time. Sensing-assisted communication therefore requires observations that can be associated with channel states at corresponding locations and times, especially as vehicle motion changes links between line-of-sight (LoS) and non-line-of-sight (NLoS) conditions~\cite{yang2023environmentsemantics}.

Different environmental sensors describe complementary aspects of the propagation environment. Red-green-blue (RGB) images provide visual and semantic context, light detection and ranging (LiDAR) point clouds capture three-dimensional (3D) geometry, and radio detection and ranging (Radar) measurements provide range and motion information. These observations have supported vision-assisted beamforming, millimeter-wave (mmWave) blockage prediction, and global positioning system (GPS)-aided vehicular beam tracking~\cite{ahn2023visionbeamforming,yang2023environmentsemantics,bian2024lidargpsbeamtracking}. Multi-modal feature fusion and structured scene representations subsequently map this environmental information to beam-management decisions~\cite{zhang2026moeisac,bian2026gsbf,bian2026virtualbasestationalignment,bian2026virtualbasestationmanagement}. When the direct path is blocked, an RIS makes the reflected channel configurable through its phase profile. Camera-assisted RIS methods therefore require visual observations paired with phase-dependent channel responses~\cite{jiang2023risvisionbeamselection}.

The required sensing--wireless association is difficult to obtain outside an instrumented measurement campaign. DeepSense~6G provides measurements from such scenarios~\cite{alkhateeb2023deepsense}, but extending coverage to new sites, trajectories, and radio configurations requires further acquisition campaigns.

Synthetic datasets provide controllable sensing observations and wireless supervision. DeepMIMO provides reproducible ray-traced channels~\cite{alkhateeb2019deepmimogenericdeeplearning}, while Sionna provides wireless simulation tools~\cite{hoydis2023sionna}, and Multimodal-Wireless generates synchronized sensing and channel data across predefined virtual towns~\cite{mao2026multimodalwireless}. For a user-specified site, the remaining challenge is to reconcile geographic records into a navigable 3D scene while preserving the correspondence between sensing observations and channel responses across controlled trajectories, dynamic blockage, and configurable reflected propagation.

To address this challenge, we propose a site-grounded multi-modal ISAC data generation (SiGMA) framework. SiGMA reconstructs a specified geographic environment and derives sensing observations and wireless supervision from corresponding scene states along configured vehicle trajectories. This design supports LoS, NLoS, and changing blockage conditions, as well as phase-configurable reflected channels when an RIS is specified. The main contributions are as follows:
\begin{itemize}
\item \textbf{Site-Grounded Urban Scene Construction:} We reconcile OpenStreetMap (OSM) and Overture Maps under spatial consistency constraints and construct georeferenced 3D urban environments through procedural modeling for sensing and wireless simulation.
\item \textbf{Shared-State Multi-Modal Data Generation:} We associate sensing observations with wireless channel data through a common scene geometry and dynamic actor state. We further provide configurable frequency-modulated continuous-wave (FMCW) Radar in-phase and quadrature (IQ) synthesis for receiver-side processing.
\item \textbf{Propagation-Controlled Communication Studies:} We use predefined mobility to sample LoS, NLoS, and changing blockage conditions. For RIS-assisted links, we generate component and cascaded channels that connect phase configurations to the same scene states, enabling zero-shot beam prediction transfer and camera-assisted RIS beam-selection studies.
\end{itemize}

\section{SiGMA Dataset Architecture and Pipeline}
\label{sec:scope}

In this section, we define the dataset contract that preserves correspondence between sensing and wireless records. Fig.~\ref{fig:overview} illustrates how SiGMA converts a deployment specification into physically associated records. The configuration defines a geographic region, roadside unit (RSU) and sensor poses, predefined vehicle routes, and radio parameters. Three stages establish the static environment, instantiate dynamic world states, and synthesize the corresponding propagation responses.

\begin{itemize}
\item \textbf{Stage I: Site-Grounded Map Construction.} Geographic sources are reconciled into a georeferenced 3D scene with drivable roads and semantic assets for traffic and sensor simulation.
\item \textbf{Stage II: Multi-Modal Sensing Acquisition.} Predefined routes and dynamic traffic generate time-varying visibility and blockage states alongside CARLA sensing observations.
\item \textbf{Stage III: Wireless Data Synthesis.} Ray tracing maps corresponding scene states to wireless channels and Radar responses. Sionna provides Radar paths and the channel impulse response (CIR) for FMCW IQ synthesis. When an RIS is configured, its component channels form phase-dependent cascaded channels.
\end{itemize}

The resulting dataset associates modalities through site, trajectory, and frame identifiers. Modality-specific sample indices preserve acquisition-rate differences, while geographic provenance and deployment metadata support repeated generation under controlled configurations.

\section{SiGMA Data-Generation Pipeline}
\label{sec:gen}

In this section, we describe how geographic reconstruction, road-constrained mobility, and propagation synthesis implement the shared-state data contract. The sensing and propagation records are derived from the same scene geometry and time-indexed actor states.

\subsection{Site-Grounded Map Construction}
\label{sec:geo}
\begin{figure*}[t!]
\centering
\includegraphics[width=.99\textwidth]{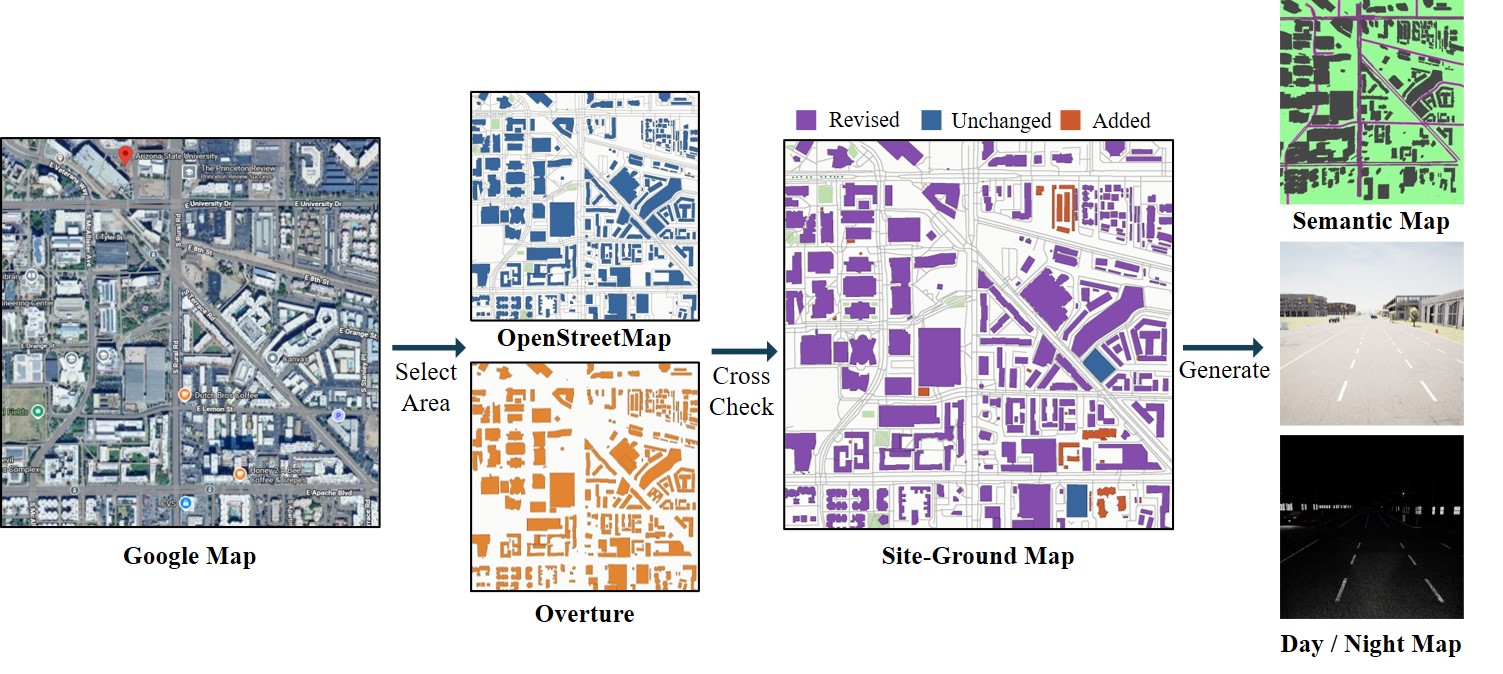}
\caption{Site-grounded urban reconstruction. OSM roads and building information are reconciled with complementary Overture footprints before procedural scene generation. The resulting scene is represented through semantic and rendered urban views.}
\label{fig:site}
\end{figure*}

Fig.~\ref{fig:site} illustrates the site-grounded urban reconstruction process. SiGMA uses OSM for the primary road network and baseline building footprints~\cite{mordechai2008osm}. Overture Maps supplies candidate footprints where OSM building coverage is incomplete. After projection into a common geographic reference, OSM footprints take precedence and geometric overlap and proximity checks screen candidate additions for conflicts with buildings, road corridors, water bodies, and map boundaries. Accepted features retain their source attribution, while available height attributes can be incorporated through optional enrichment. 

The road network is converted through CARLA's OSM-to-OpenDRIVE interface into a simulator-compatible representation, with the projected origin retained for infrastructure placement and navigation. Building footprints and available height attributes guide procedural construction of the 3D geometry and facades. Roads, vegetation, terrain, and roadside assets complete the urban environment in Unreal Engine. Semantic categories support CARLA rendering, while electromagnetic material assignments support wireless propagation.

Post-construction checks verify road connectivity, route reachability, geometric consistency, and the semantic categories required by downstream sensing and propagation models.

\subsection{Multi-Modal Sensing Acquisition}

Road-constrained traffic creates time-varying sensing and propagation conditions. Each communication-vehicle trajectory is specified by a start, destination, and ordered waypoints on the road network. Feasible paths connect successive waypoints according to road topology and reachability constraints. Background traffic and route geometry create dynamic obstructions and vary visibility, producing LoS/NLoS conditions and transitions along each trajectory.

CARLA advances the simulation in synchronous steps and records actor states alongside sensor frames. At instant $t$, the shared world state $\mathbf{s}_t$ contains the identities, positions, orientations, dimensions, and velocities of relevant actors. Sensors acquire observations at configured intervals aligned with the simulation time step, either within the RSU camera field of view (FoV) or along a predefined route interval.

RGB, depth, and LiDAR observations are rendered from configured sensor poses, while GPS and inertial measurement unit (IMU) records provide positioning and motion information. Frame identifiers link each observation to $\mathbf{s}_t$ while preserving modality-specific sampling intervals. A consistent CARLA-to-radio coordinate transformation, $(x,y,z)\mapsto(x,-y,z)$, together with the corresponding orientation transformation, maps the static scene and frame-specific actor poses into Blender and Sionna. This shared physical reference maintains spatial correspondence between sensing and propagation data.

\subsection{Wireless and Radar Data Generation}

\subsubsection{Ray-Traced Wireless Channels}
\label{sec:wireless}

The static environment and time-varying actor geometry are transferred through Blender into Sionna-compatible scenes for ray tracing~\cite{hoydis2023sionna}. Configured radio materials, antenna arrays, carrier frequency, and interaction mechanisms determine the complex path gains, delays, and propagation directions. The frequency-domain channel response for receive element $m$ and transmit element $n$ follows the path sum,
\begin{equation}
\label{eq:wideband}
[\mathbf{H}_t(f)]_{m,n}
=\sum_{\ell\in\mathcal{P}_{m,n,t}}
a_{m,n,\ell,t}e^{-j2\pi f\tau_{m,n,\ell,t}},
\end{equation}
where $f$ is the frequency offset from the carrier, $\mathcal{P}_{m,n,t}$ is the set of propagation paths, and $a_{m,n,\ell,t}$ and $\tau_{m,n,\ell,t}$ denote the complex path gain and propagation delay. Complex path gains include carrier-frequency phase, and the evolution of these gains across snapshots captures motion-induced Doppler. Far-field channels use synthetic-array modeling with plane-wave array responses. Near-field channels use explicit per-element ray tracing to capture element-dependent propagation. Path types indicate direct-path visibility, and path powers characterize dominant propagation components along vehicle trajectories.

For RIS-assisted communication, ray tracing provides the RSU-to-RIS and RIS-to-user component channels. Their phase-weighted product determines the RIS cascaded channel for a selected phase configuration. Let $\mathcal{Q}_{\Phi}=\{\boldsymbol{\phi}_q\}_{q=1}^{Q_{\Phi}}$ denote the RIS phase codebook, with candidate reflection matrices $\boldsymbol{\Phi}_q=\operatorname{diag}(\boldsymbol{\phi}_q)$. The effective RSU--user channel combines the direct and cascaded terms,
\begin{equation}
\label{eq:ris}
\mathbf{H}_{\text{eff},t}(f)
=\mathbf{H}_{\text{SU},t}(f)
+\mathbf{H}_{\text{RU},t}(f)\boldsymbol{\Phi}_t\mathbf{H}_{\text{SR},t}(f),
\end{equation}
where $\mathbf{H}_{\text{SU}}$, $\mathbf{H}_{\text{RU}}$, and $\mathbf{H}_{\text{SR}}$ denote the direct RSU--user, RIS--user, and RSU--RIS channels, respectively. At snapshot $t$, $\boldsymbol{\Phi}_t=\boldsymbol{\Phi}_{q_t}$ for selected codeword index $q_t$. Retaining the component channels permits phase configurations to be evaluated and optimized under an unchanged scene state.

\subsubsection{FMCW Radar Signal Synthesis}
\label{sec:radar}

\begin{figure}[t]
\centering
\includegraphics[width=.99\columnwidth]{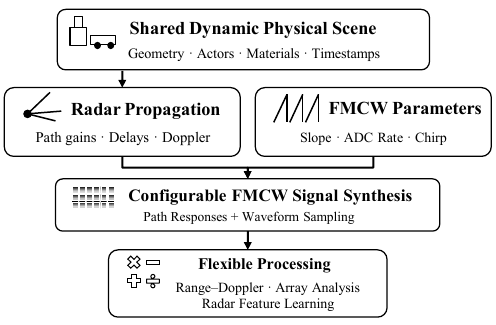}
\caption{FMCW Radar signal synthesis in SiGMA. Ray-traced Radar channel responses and configured FMCW parameters generate complex IQ samples associated with the shared dynamic scene for receiver-side processing.}
\label{fig:fmcw_radar}
\end{figure}

Fig.~\ref{fig:fmcw_radar} illustrates FMCW Radar data generation from the shared scene. Vehicle meshes determine target geometry and visibility, and a 3GPP TR 38.901 multi-scattering-point Radar cross section (RCS) model supplies target scattering responses. Sionna ray tracing provides complex path gains, absolute delays, and Doppler shifts through the scattering centers. Optional background paths represent environmental multipath. These ray-traced responses form the Radar CIR used for waveform synthesis. Each deployment specifies a monostatic transmitter and receiving array, whose carrier-frequency and element-dependent phases are retained in the CIR coefficients.

SiGMA synthesizes complex FMCW IQ samples from the Radar CIR using the configured chirp slope, analog-to-digital converter (ADC) sampling rate, chirp repetition interval, and receiver-array geometry. Propagation delay maps to a beat-frequency contribution through the chirp slope, while Doppler governs phase evolution across fast and slow time. Path gains and delays are held constant over each coherent processing interval. The resulting dechirped IQ tensor $\mathbf{Y}_t\in\mathbb{C}^{N_{\text{rx}}\times N_{\text{adc}}\times N_{\text{chirps}}}$ supports range--Doppler and angular processing. Stored path parameters permit different FMCW waveforms to be synthesized for the same scene snapshot, allowing receiver-side effects of waveform choices to be examined.

\section{Dataset Visualization and Representative Applications}
\label{sec:apps}

This section evaluates the real-world transferability of SiGMA-generated data through zero-shot beam prediction on DeepSense 6G measurements. A camera-assisted RIS case study further illustrates the framework's flexibility in generating sensing-associated supervision for configurable reflected propagation.

\subsection{Scenario-Based Dataset Visualization}

\begin{figure*}[t]
\centering
\includegraphics[width=.99\textwidth]{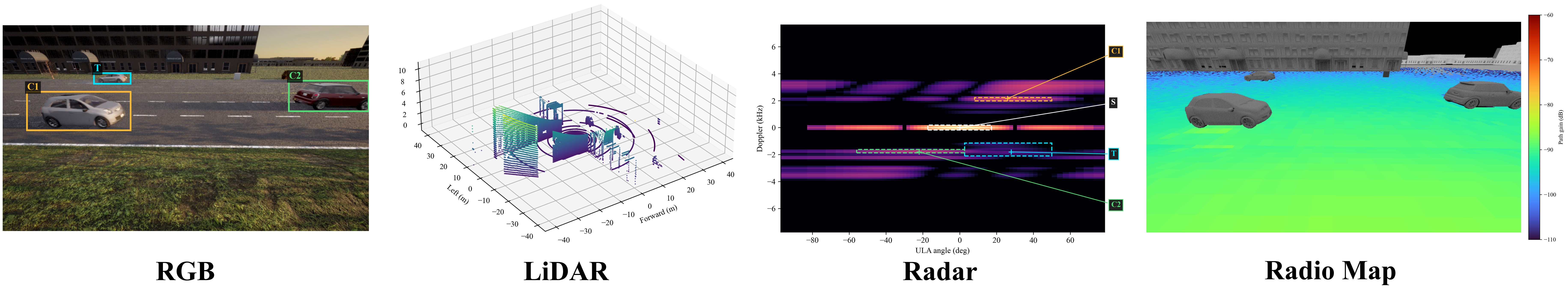}
\caption{Representative SiGMA outputs for the site corresponding to DeepSense 6G Scenario 3. The views include an annotated RGB frame, a LiDAR point cloud, an FMCW Radar Doppler--angle spectrum, and a radio map of spatial path gain.}
\label{fig:modalities}
\end{figure*}

Fig.~\ref{fig:modalities} illustrates aligned observations generated for the geographic site corresponding to DeepSense 6G Scenario 3~\cite{alkhateeb2023deepsense}. The RGB frame identifies the target vehicle T and nearby vehicles C1 and C2, while the LiDAR point cloud captures surrounding geometry and partial target occlusion. A Doppler--angle spectrum derived from the generated IQ samples represents the FMCW Radar response, and its associated path and actor information links that response to vehicle motion. The radio map gives the spatial distribution of path gain in the same environment. Together, these outputs demonstrate the physical association between sensing observations and wireless responses through a shared scene state.

\subsection{Use Case I: Site-Grounded Beam Prediction Transfer}
\begin{figure}[t]
\centering
\subfloat[Beam prediction scenario]{\includegraphics[width=.48\columnwidth]{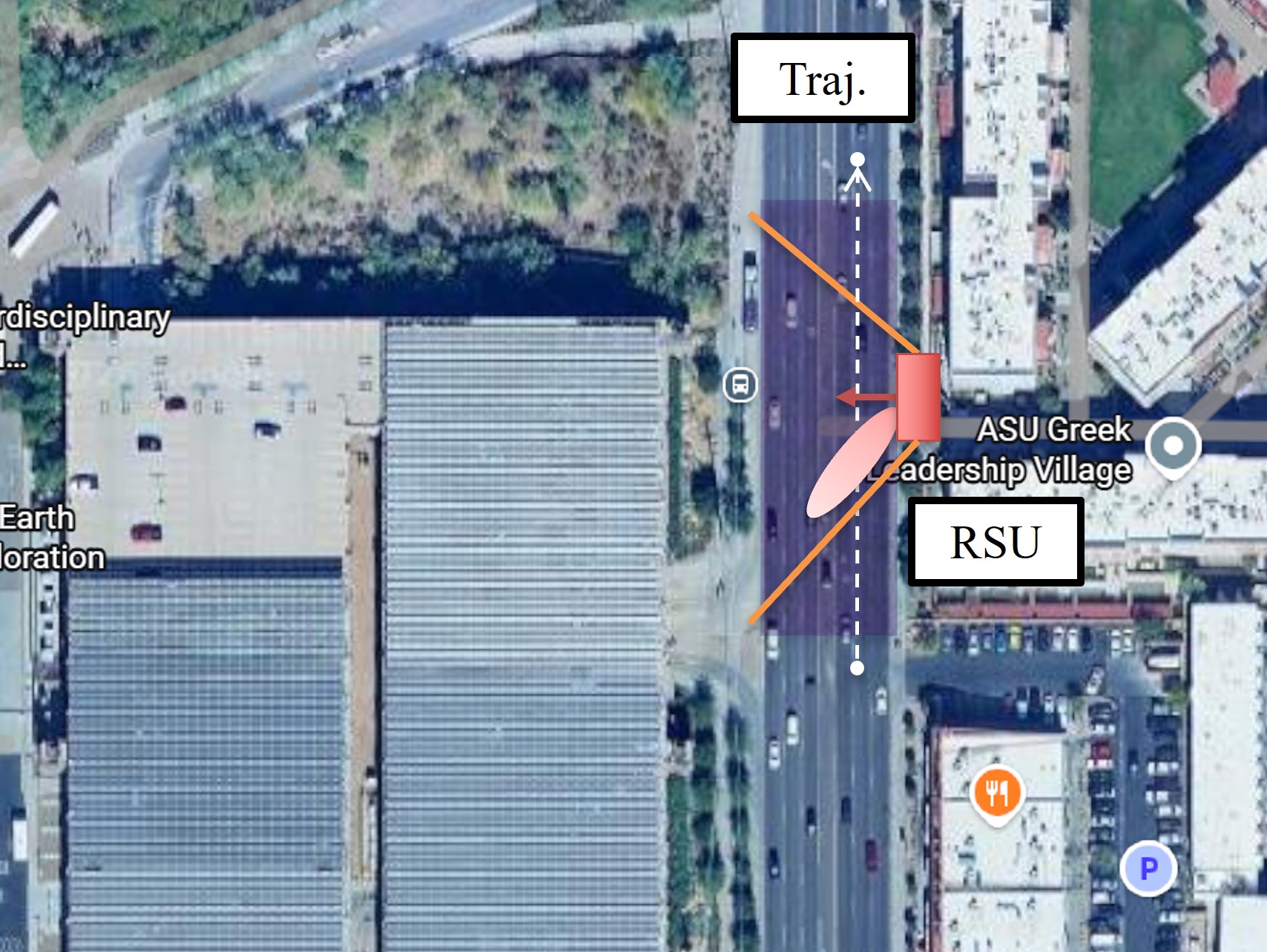}\label{fig:sim2real_scene}}\hfill
\subfloat[RIS-assisted scenario]{\includegraphics[width=.48\columnwidth]{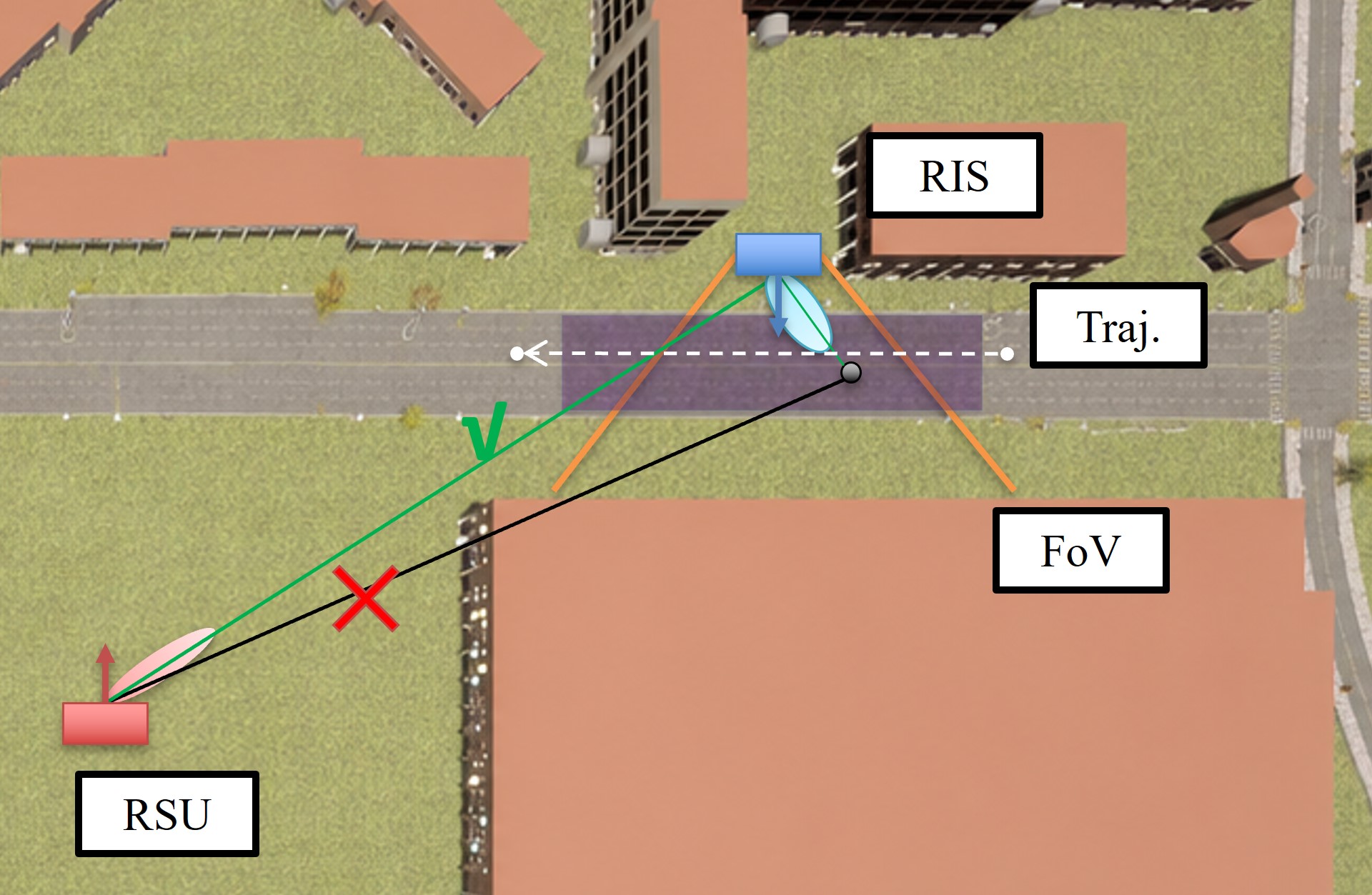}\label{fig:ris_scene}}
\caption{Deployment layouts for the two communication applications. (a) shows the RSU, communication vehicle, and camera FoV. (b) illustrates an RIS-assisted link with an obstructed direct path.}
\label{fig:use_case_scenes}
\end{figure}

Use Case I tests whether site-grounded synthetic data improve zero-shot beam prediction on the real DeepSense 6G Scenario 3 test set. We compare Camera-Assisted GPS Association trained with SiGMA data and with an OSM-derived dataset~\cite{alkhateeb2023deepsense}. Both variants use the same camera-assisted architecture and GPS-assisted association and are evaluated on 1,487 frames from 52 trajectories. A Faster R-CNN detector with a MobileNetV3 backbone extracts vehicle boxes from RSU images, and the normalized box center $c_x$ is passed to a multilayer perceptron (MLP) for beam ranking~\cite{ren2016fasterrcnnrealtimeobject,howard2019searchingmobilenetv3}. The SiGMA training set contains 3,341 frames from 50 trajectories. SiGMA combines OSM-Overture geographic reconciliation with automated 3D reconstruction, procedural facade modeling, and texture rendering. The OSM-derived baseline constructs buildings from OSM footprints as simplified, untextured volumes. Fig.~\subref{fig:sim2real_scene} shows the RSU deployment.

We consider an RSU equipped with $N_{\text{RSU}}$ antennas and a transmit codebook $\mathcal F=\{\mathbf f_1,\ldots,\mathbf f_Q\}$. The 32-beam codebook selects every second beam from the original 64-beam codebook, with optimal labels recomputed over the selected beams. Given channel $\mathbf h_t$, the codebook-optimal beam index maximizes the received power,
\begin{equation}
\label{eq:beam}
q_t^\star=\arg\max_{1\le q\le Q}
|\mathbf h_t^{\mathsf H}\mathbf f_q|^2.
\end{equation}
During training, SiGMA channel responses provide optimal-beam labels for MLP supervision. At test time, real camera observations and GPS-assisted target association provide the model input for ranking candidate beams. Measured per-beam powers in DeepSense 6G determine the reference beam labels and the power recovered by each predicted candidate set, yielding the zero-shot evaluation.

\subsection{Use Case II: Camera-Assisted RIS Beam Selection}

Site geometry and vehicle position determine direct-link visibility. When the direct path is blocked, RIS-assisted reflected propagation can provide a configurable alternative through phase selection. This use case pairs camera observations with the RSU-to-RIS and RIS-to-vehicle component channels from corresponding scene states to supervise phase selection. Samples include LoS and NLoS conditions, and Fig.~\subref{fig:ris_scene} shows a representative blocked-direct-link deployment. The panel has $32\times32=1{,}024$ elements. Phase codebooks with $Q_{\Phi}\in\{256,1{,}024\}$ cover $16\times16$ and $32\times32$ angular directions in the camera field of view, respectively, with fixed incident-wave compensation.

A COCO-pretrained Faster R-CNN with a MobileNetV3 backbone extracts vehicle boxes from RIS camera images. One known target box initializes tracking per trajectory, and the MLP ranks configurations using normalized box information~\cite{ren2016fasterrcnnrealtimeobject,howard2019searchingmobilenetv3}. We evaluate RIS beam selection on 2,616 simulated frames from 50 trajectories using three-fold group-based cross-validation. Geometrically similar trajectories are grouped to assess generalization across held-out routes.

The RSU is equipped with 16 antennas and uses a fixed, unit-norm beam $\mathbf f_{\text{RSU}}$ directed toward the RIS. The vehicle has one receive antenna. Let $\mathbf{h}_{\text{RU},t}\in\mathbb{C}^{N_{\text{RIS}}\times1}$ be the RIS-to-vehicle channel vector and $\mathbf{H}_{\text{SR},t}\in\mathbb{C}^{N_{\text{RIS}}\times16}$ the RSU-to-RIS channel. The RIS-only label selects the configuration with maximum cascaded-link power under this beam,
\begin{equation}
\label{eq:ris_beam}
q_t^\star=\arg\max_{1\le q\le Q_{\Phi}}
\left|\mathbf{h}_{\text{RU},t}^{\mathsf H}
\boldsymbol{\Phi}_q\mathbf{H}_{\text{SR},t}\mathbf{f}_{\text{RSU}}\right|^2.
\end{equation}
Following prior camera-assisted RIS beam-selection work~\cite{jiang2023risvisionbeamselection}, the classifier ranks candidate phase configurations for each associated vehicle frame. Candidate powers and labels use the RIS cascaded term in \eqref{eq:ris}, while the general effective-channel model also includes the direct channel.

\subsection{Evaluation Metrics and Results}
\label{sec:results}
For test frame $t$, let $P_{t,q}$ denote received power, $\mathcal C$ the full codebook, and $\mathcal S_{t,k}$ the selected Top-$k$ set. In Use Case I, $q$ indexes beams and $P_{t,q}$ denotes measured beamforming power. In Use Case II, $q$ indexes RIS configurations and $P_{t,q}$ denotes RIS cascaded-link power. Across $T$ frames, normalized gain measures the best selected power relative to the full-codebook optimum,
\begin{equation}
\label{eq:normalized_gain}
\mathrm{NG}@k=\frac{1}{T}\sum_{t=1}^{T}
\frac{\max_{q\in\mathcal S_{t,k}}P_{t,q}}
{\max_{q\in\mathcal C}P_{t,q}},
\end{equation}
where $T$ is the number of test frames. A failed detection, target association, or prediction contributes zero power to the numerator. Top-$k$ accuracy measures whether the codebook-optimal candidate belongs to $\mathcal S_{t,k}$. Each codebook uses its own full-scan optimum as the normalization reference.

\subsubsection{Zero-Shot Beam Prediction Transfer}
\begin{figure}[t]
\centering
\subfloat[Codebook size $Q=32$]{\includegraphics[width=.9\columnwidth]{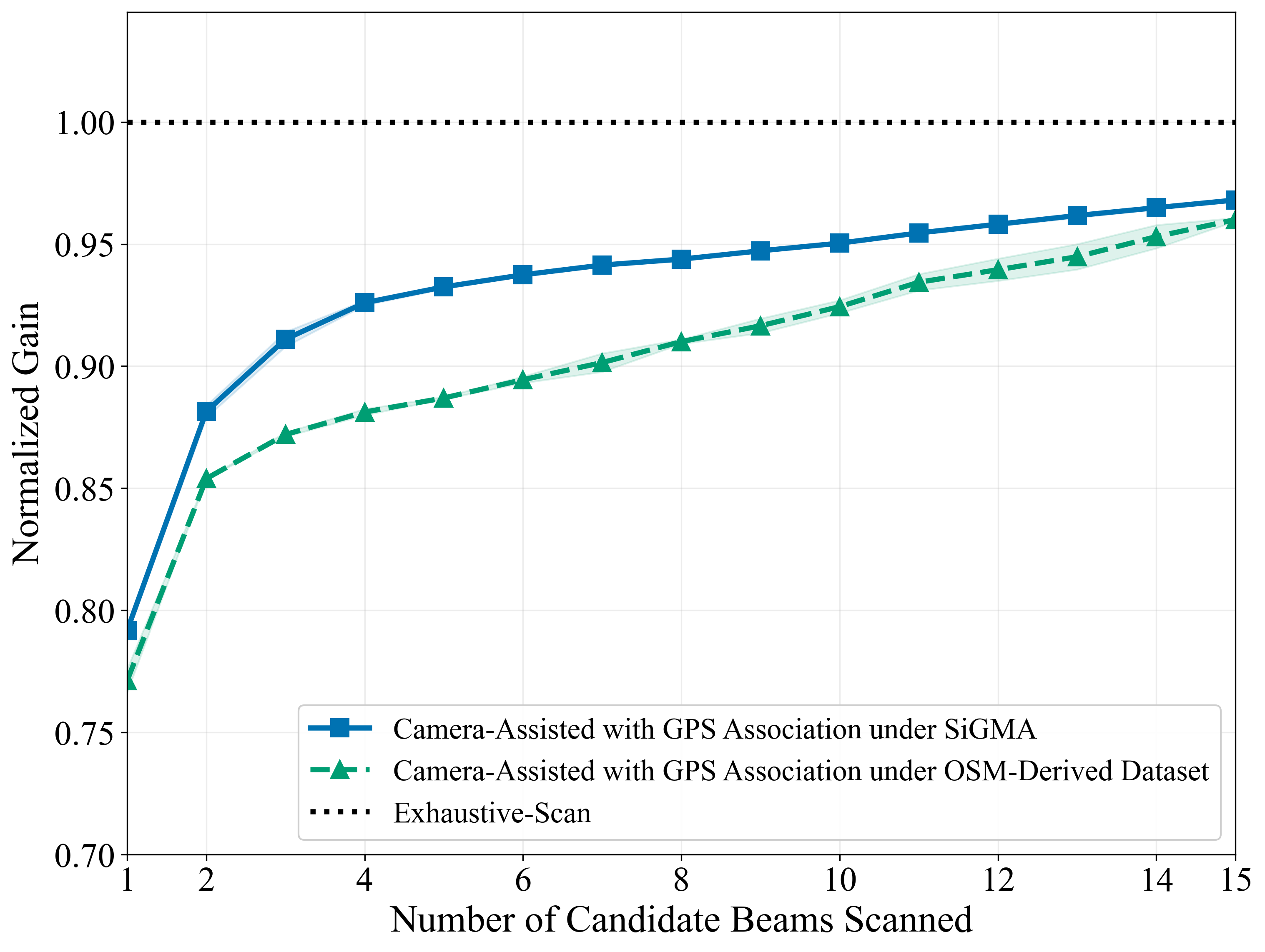}\label{fig:beam_transfer_32}}\hfill
\subfloat[Codebook size $Q=64$]{\includegraphics[width=.9\columnwidth]{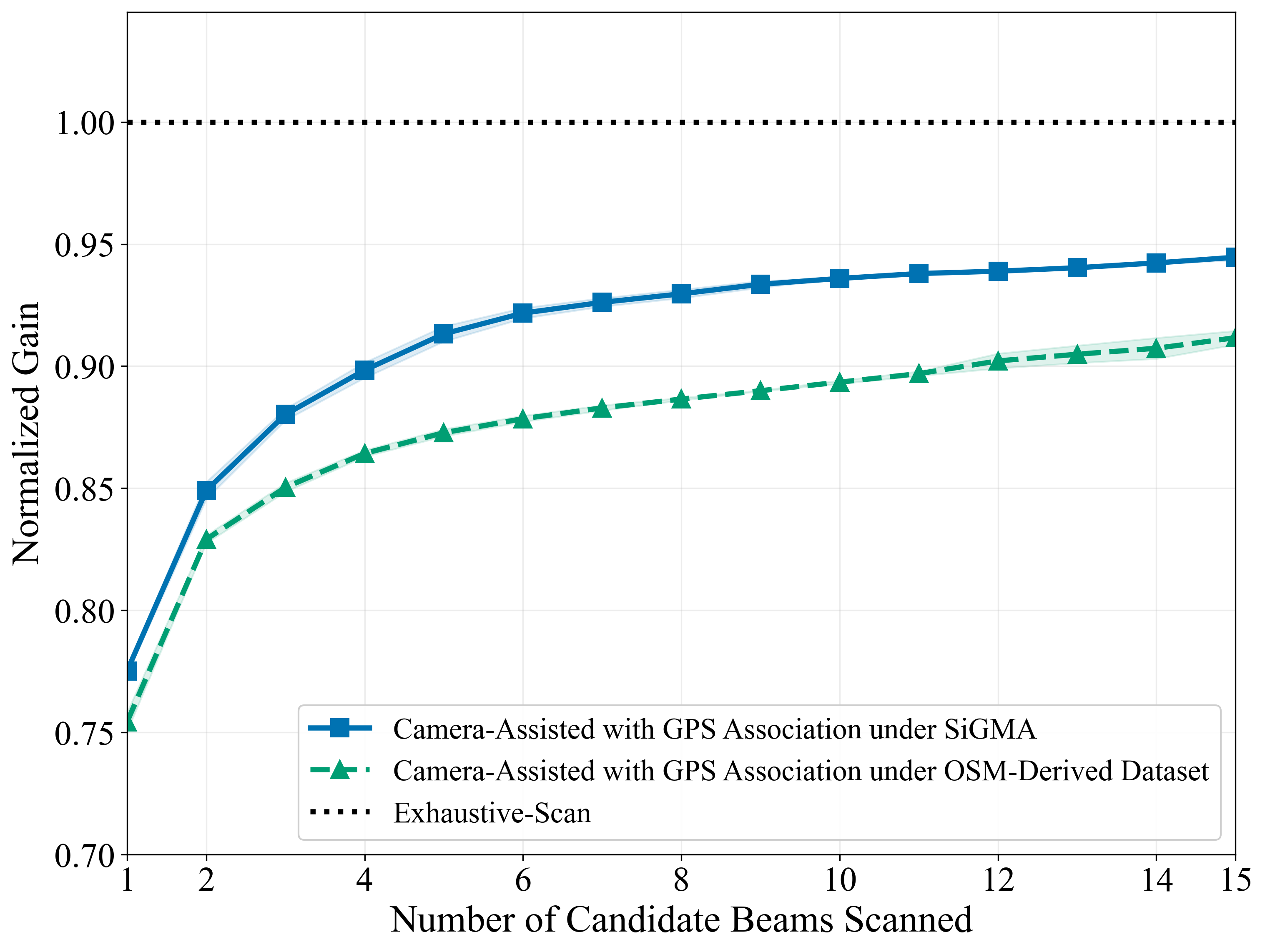}\label{fig:beam_transfer_64}}
\caption{The SiGMA-trained predictor and OSM-derived baseline on real DeepSense 6G Scenario 3 measurements. Panels (a) and (b) show normalized gain in beamforming power for codebook sizes $Q=32$ and $Q=64$, respectively. The dotted Exhaustive-Scan curve marks the full-codebook reference.}
\label{fig:beam_transfer}
\end{figure}

Fig.~\ref{fig:beam_transfer} compares zero-shot beam prediction under different scan budgets. For \(Q=32\) and \(k=5\), the SiGMA-trained predictor achieves approximately 0.93 normalized gain, compared with 0.89 for the OSM-derived baseline.  The dotted Exhaustive-Scan curve marks the normalized-gain upper bound under exhaustive beam scanning. For \(Q=64\) and \(k=5\), the SiGMA-trained predictor achieves approximately 0.91 normalized gain, compared with 0.87 for the OSM-derived baseline. The improvement is consistent with the closer correspondence between reconstructed scene geometry, visual observations, and wireless propagation. Both methods approach the full-codebook reference as the scanning budget increases.

\subsubsection{Camera-Assisted RIS Beam Selection}
\begin{figure}[t]
\centering
\subfloat[$Q_{\Phi}=256$ candidates]{\includegraphics[width=.9\columnwidth]{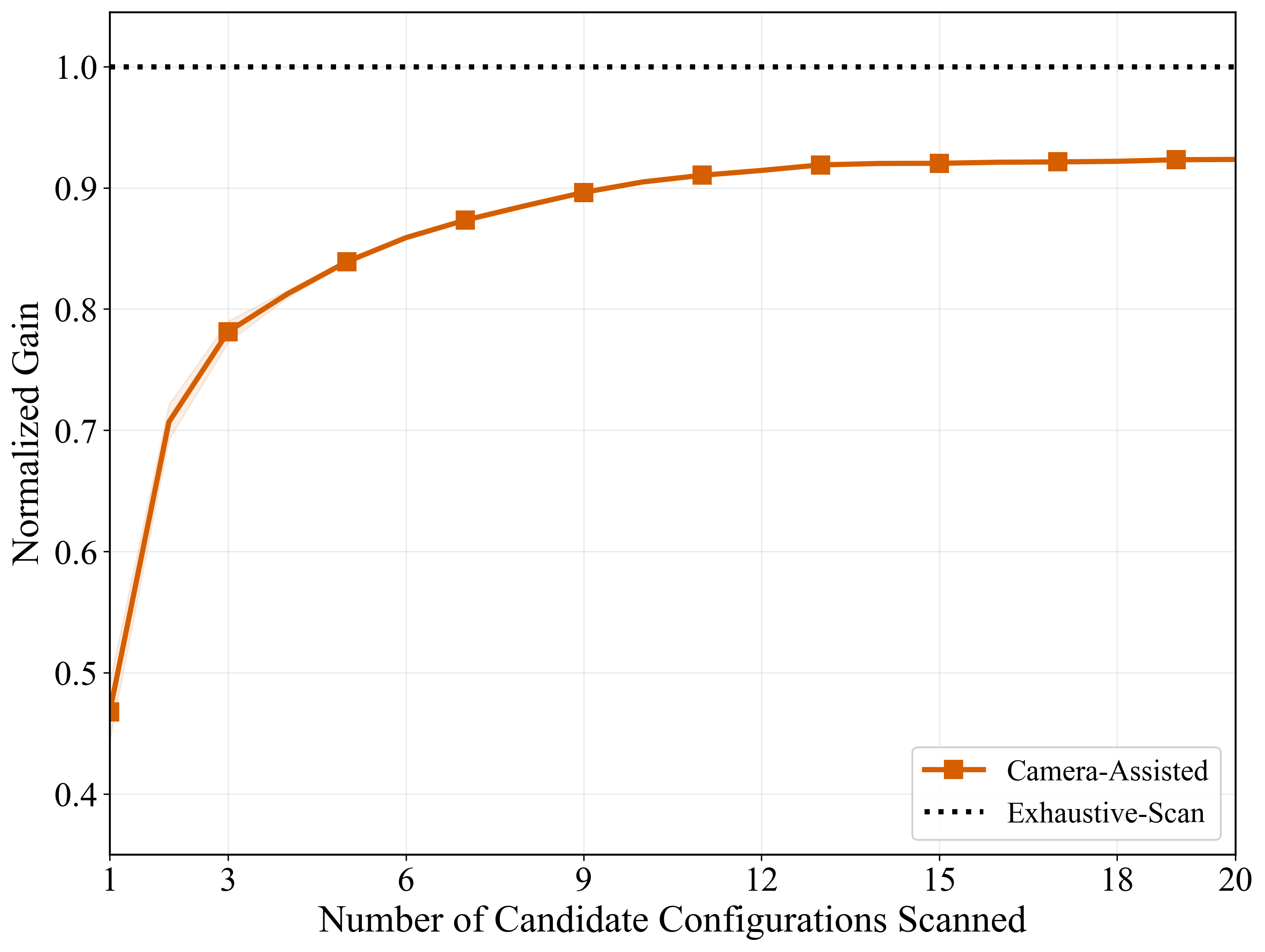}\label{fig:ris_selection_256}}\hfill
\subfloat[$Q_{\Phi}=1,024$ candidates]{\includegraphics[width=.9\columnwidth]{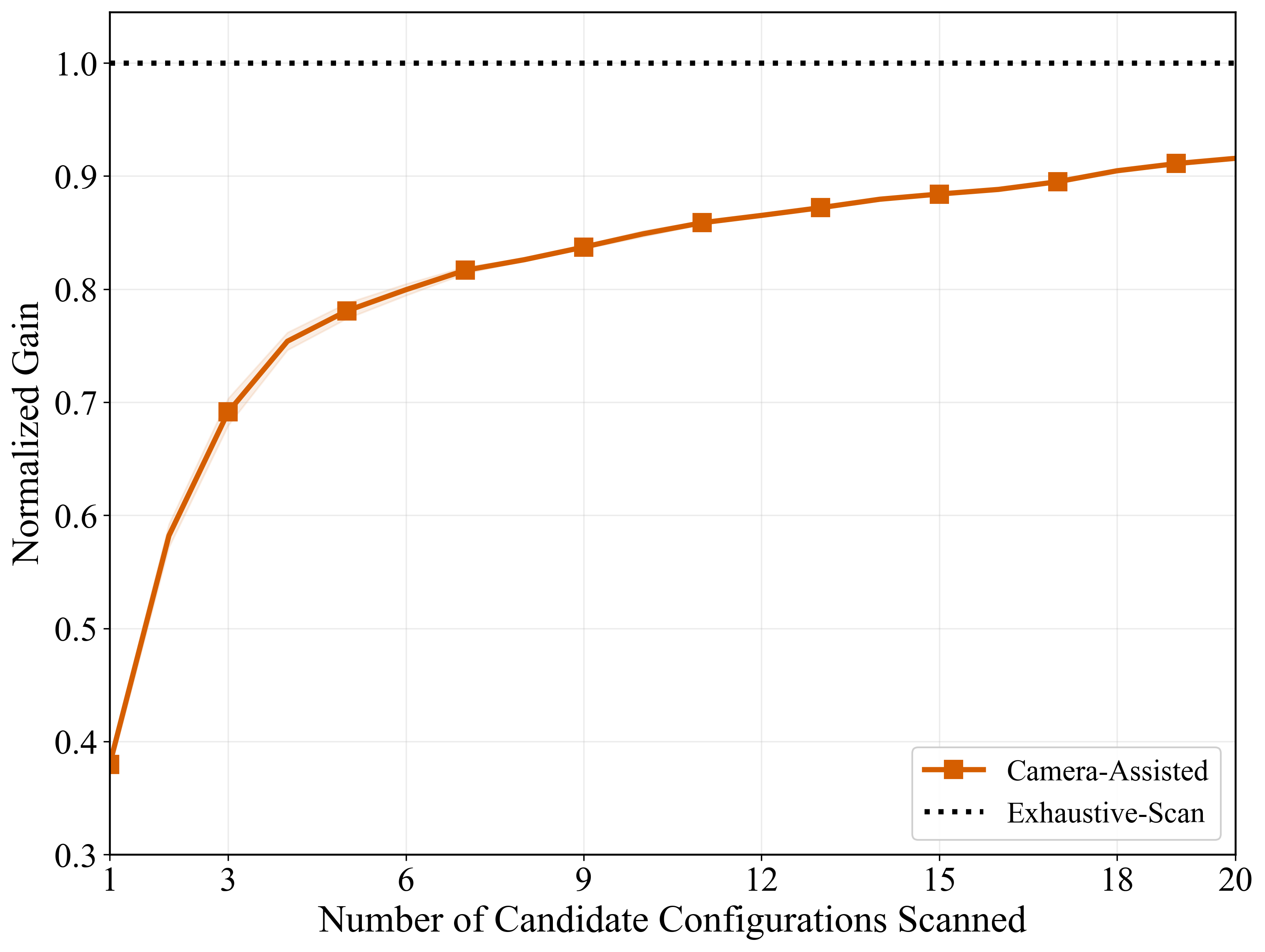}\label{fig:ris_selection_1024}}
\caption{Normalized gain in RIS cascaded-link power versus the number of scanned phase configurations. Panels (a) and (b) compare $Q_{\Phi}=256$ and $Q_{\Phi}=1,024$ candidate codebooks on the same 1,024-element RIS panel.}
\label{fig:ris_selection}
\end{figure}

Fig.~\ref{fig:ris_selection} plots normalized gain in RIS cascaded-link power against scan budget. At $k=5$, gains are approximately 0.84 and 0.78 for $Q_{\Phi}=256$ and $1{,}024$, respectively, and both approach 0.92 at $k=20$. The larger codebook recovers a smaller fraction at $k=5$, with the gap narrowing as scans increase. Top-5 accuracies are 81.63\% and 73.51\%.

\section{Conclusion}
In this paper, we presented SiGMA, a framework for site-grounded multi-modal ISAC data generation. Geographic reconstruction and shared scene states enable synchronized sensing and wireless supervision across configurable mobility and propagation conditions. Improved zero-shot beam prediction on real DeepSense 6G measurements indicates the potential to reduce real-world data acquisition requirements. The RIS case study further illustrates support for configurable reflected propagation. These capabilities facilitate geographically grounded dataset generation and provide a foundation for future AI-native ISAC research.

\bibliographystyle{IEEEtranAutoAbbr}
\bibliography{references}
\end{document}